\documentclass[aps,twocolumn,groupedaddress]{revtex4}

\usepackage{graphics,epsfig}

\begin{document}

\title{Lepton Flavor Violation in the Two Higgs Doublet Model type III}
\author{Rodolfo A. Diaz}
\email[radiaz@ciencias.ciencias.unal.edu.co]{}
\author{R. Martinez}
\email[romart@ciencias.ciencias.unal.edu.co]{}                        
\author{J-Alexis Rodriguez} 
\email[alexro@ciencias.ciencias.unal.edu.co]{}
\affiliation{Departamento de Fisica, Universidad Nacional de Colombia\\
Bogota, Colombia}

\begin{abstract}
We consider the Two Higgs Doublet Model (2HDM) of type III which leads to
Flavour Changing Neutral Currents (FCNC) at tree level in the leptonic
sector. In the framework of this model we can have, in principle, two
situations: the case (a) when both doublets acquire a vacuum expectation
value different from zero and the case (b) when only one of them is not
zero. In addition, we show that we can make two types of rotations for the
flavor mixing matrices which generates four types of lagrangians, with the
rotation of type I we recover the case (b) from the case (a) in the limit $%
\tan \beta \rightarrow \infty $, and with the rotation of type II we obtain
the case (b) from (a) in the limit $\tan \beta \rightarrow 0.\;$Moreover,
two of the four possible lagrangians correspond to the models of types I and
II plus Flavor Changing (FC) interactions. The analitical expressions of the
partial lepton number violating widths $\Gamma \left( \mu \rightarrow
eee\right) $ and $\Gamma (\mu \rightarrow e\gamma )\;$are derived for the
cases (a) and (b) and both types of rotations.$\;$In all cases these widths
go asymptotically to zero in the decoupling limit for all Higgses. We
present from our analysis upper bounds for the flavour changing transition $%
\mu \rightarrow e,$ and we show that such bounds are sensitive to the VEV
structure and the type of rotation utilized.
\end{abstract}

\maketitle


\preprint{UNAL,GFTAE-3-00}








 


\section{Introduction}

Flavor Changing Neutral Currents (FCNC) are forbidden at tree level in the
Standard Model (SM). However, they could be present at one loop level as in
the case of $b\rightarrow s\gamma $ \cite{bsg}, $K^{0}\rightarrow \mu
^{+}\mu ^{-}$ \cite{kmm}, $K^{0}-\overline{K}^{0}$ \cite{koko}, $%
t\rightarrow c\gamma $ \cite{tcg} etc. In general, many extensions of the SM
permit, however FCNC at tree level. The introduction of new representations
of fermions different from doublets produce them by means of the Z-coupling 
\cite{2}. In addition, they are generated at tree level in the Yukawa sector
by adding a second doublet to the SM \cite{wolf}. Such couplings also appear
in SUSY theories without R-parity \cite{R1}. Theories with FCNC were
previously considered unatractive because they were strongly constrained
experimentally, especially due to the small $K_{L}-K_{S}\;$mass difference.
Nevertheless, nowadays it is hoped to observe such physical processes in
laboratory, as a result many theories were proposed (see above).

Owing to the continuous improvements in experimental accuracies, Lepton
Flavor Violation (LFV) has become a very important possible source of new
physics. Experiments to search directly for LFV have been performed for many
years, all with null results so far. Experimental limits have resulted from
searches for $K_L^0 \to \mu^+ e^-$ \cite{arisaka}, $K_L^0 \to \pi^0 \mu^+
e^- $ \cite{plb}, $K^+ \to \pi^+ \mu^+ e^-$ \cite{lee}, $\mu^+ \to e^+
\gamma $ \cite{bolton}, $\mu^+ \to e^+ e^+ e^-$ \cite{bell} and $\mu^- N \to
e^-N$ \cite{doh}.

There are several mechanisms to avoid FCNC at tree level. Glashow and
Weinberg \cite{gw} proposed a discrete symmetry in the Two Higgs Doublet
Model (2HDM) which forbids the couplings that generate such rare decays,
hence they do not appear at tree level. Another possibility is to consider
heavy exchange of scalar or pseudoscalar Higgs fields \cite{Sher91} or by
cancellation of large contributions with opposite sign. Another mechanism
was proposed by Cheng and Sher arguing that a natural value for the FC
couplings from different families should be of the order of the geometric
average of their Yukawa couplings \cite{ChengSher}.

Taking this \textit{natural} assumption and since Yukawa couplings in the SM
vary with mass, it is plausible that the same occurs for FC couplings. Hence
it is expected that FCNC involving the third generation can be larger, while
the ones involving the first generation are hoped to be small \cite{Sher91}, 
\cite{Reina}. Another clue that suggests large mixing between the second and
third generation in the charged leptonic sector, is the large mixing between
second and third generation of the neutral leptonic sector. This is
predicted by experiments with atmospheric neutrinos \cite{Fukuda}.

The increasing interest in LFV processes is due to the strong restrictions
that experiments have imposed on them. This consequently determines small
regions of parameters for new physics of any theory beyond the SM. Some
specific decays have been widely studied within the framework of
supersymmetric extensions, because in Supersymmetric theories the presence
of FCNC induced by R-parity violation generates massive neutrinos and
neutrino oscillations \cite{Kaustubh}. In recent papers the decays $\mu
\rightarrow e\gamma \;$and $\mu \rightarrow 3e\;$with polarized muons have
been examined in the context of supersymmetric grand unified theories to get
bounds in the $m_{\widetilde{e}_{R}}-\left| A_{0}\right| \;$plane \cite
{Okada}.

On the other hand, a muon collider could provide very interesting new
constraints on FCNC, for example $\mu \mu \rightarrow \mu \tau (e\tau )\;$
mediated by Higgs exchange \cite{SherCollider} which test the mixing between
the second and third generations. Additionally, the muon collider could be a
Higgs factory and it is well known that the Higgs sector is crucial for FCNC 
\cite{Workshop}. Finally, effects on the coupling of muon and tau in the
2HDM framework owing to anomalous magnetic moment of the muon could be
significantly improved by E821 experiment at Brookhaven National Laboratory 
\cite{SherCollider}.

Additionally, in the quark sector bounds on LFV come from $\Delta F=2$
processes, rare B-decays, $Z \to \overline b b$ and the $\rho$-parameter 
\cite{ARS}. Reference \cite{ARS} also explored the implications of FCNC at
tree level for $e^+e^- (\mu^+ \mu^-) \to t \overline c + \overline t c$, $t
\to c \gamma (Z,g)$, $D^0-\overline D^0$ and $B^0_s-\overline B^0_s$.
Moreover, there are other important processes involving FCNC. For instance,
the decay $B^{-}(D^-)\rightarrow K^{-}\mu ^{+}\tau ^{-}\;$ which depends on $%
\mu -\tau \;$mixing and vanishes in the SM. Hence it is very sensitive to
new physics. Another one is $B^{-}(D^-)\rightarrow K^{-}\mu ^{+}e^{-}\;$%
whose form factors have been calculated in \cite{Sher91}, \cite{nosotros}.

The simplest model which exhibits FCNC at tree level is the model with one
extra Higgs doublet, known as the two Higgs doublet model (2HDM). There are
several kinds of such models. In the model type I, one Higgs Doublet
provides masses to the up and down quarks, simultaneously. In the model type
II, one Higgs doublet gives masses to the up quarks and the other one to the
down quarks. These former two models have the discrete symmetry mentioned
above to avoid FCNC at tree level \cite{gw}. However, the discrete symmetry
is not necessary in whose case both doublets generate the masses of the
quarks of up-type and down-type, simultaneously. In the literature, the
latter is known as the model type III \cite{III}. It has been used to look
for physics beyond the SM and specifically for FCNC at tree level \cite{ARS}%
, \cite{Sher91}. In general, both doublets could acquire a vacuum
expectation value (VEV), but we can absorb one of them redefining the Higgs
fields properly. Nevertheless, we shall show that a substantial difference
arises from the case in which both doublets get the VEV, and therefore we
will study the model type III considering two cases. In the first case, the
two Higgs doublets acquire VEV (case (a)). In the second one, only one Higgs
doublet acquire VEV (case (b)). In the latter case the free parameter $\tan
\beta $\ is removed from the theory making the analysis simpler.

In section II, we describe the model and define the notation we shall use
throughout the document. In section III, we show that we can make two
kinds of rotations for the flavor mixing matrices which generates four types
of lagrangians, and that in the framework of the first rotation we arrive to
the case (b) from the case (a) in the limit $\tan \beta \rightarrow \infty $%
, while with the second rotation we obtain (b) from (a) in the limit $\tan
\beta \rightarrow 0.\;$Furthermore, we find that two of the four possible
lagrangians correspond to the models of types I and II plus Flavor Changing
(FC) interactions.

In section IV we get bounds on LFV in the 2HDM type III based on the decays $%
\mu \rightarrow e\gamma $ and $\mu \rightarrow eee$. Such decays are
examined in the framework of both cases (a) and (b) according to the
classification made above, and with both types of rotations. We find that
such constraints depend on whether we use cases (a) or (b) and on what kind
of rotation is utilized.

\section{The Model}

The 2HDM type III is an extension of the SM plus a new Higgs doublet and
three new Yukawa couplings in the quark and leptonic sectors. The mass terms
for the up-type or down-type sectors depends on two matrices or two Yukawa
couplings. The rotation of the quarks and leptons allows us to diagonalize
one of the matrices but not both simultaneously, so one of the Yukawa
couplings remains non-diagonal, generating the FCNC at tree level.

The Yukawa's Lagrangian is as follow 
\begin{eqnarray}
-\pounds _{Y} &=&\eta _{ij}^{U,0}\overline{Q}_{iL}^{0}\widetilde{\Phi }%
_{1}U_{jR}^{0}+\eta _{ij}^{D,0}\overline{Q}_{iL}^{0}\Phi _{1}D_{jR}^{0}+\eta
_{ij}^{E,0}\overline{l}_{iL}^{0}\Phi _{1}E_{jR}^{0}  \label{Yukawa} \\
&+&\xi _{ij}^{U,0}\overline{Q}_{iL}^{0}\widetilde{\Phi }_{2}U_{jR}^{0}+\xi
_{ij}^{D,0}\overline{Q}_{iL}^{0}\Phi _{2}D_{jR}^{0}+\xi _{ij}^{E,0}\overline{%
l}_{iL}^{0}\Phi _{2}E_{jR}^{0}+h.c.  \nonumber
\end{eqnarray}
where $\Phi _{1,2}\;$are the Higgs doublets,$\;\eta _{ij}^{0}\;$and $\xi
_{ij}^{0}\;$ are non-diagonal $3\times 3\;$non-dimensional matrices and $i$, 
$j$ are family indices. $D\;$refers to the three down quarks $D\equiv \left(
d,s,b\right) ^{T},\;U\;$refers to the three up quarks $U\equiv \left(
u,c,t\right) ^{T}\;$and $E\;$to the three charged leptons. The superscript $%
0 $ indicates that the fields are not mass eigenstates yet. In the so-called
model type I, the discrete symmetry forbids the terms proportional to $\xi
_{ij}^{0},\;$meanwhile in the model type II the same symmetry forbids terms
proportional to $\xi _{ij}^{U,0},\;\eta _{ij}^{D,0},\eta _{ij}^{E,0}.$

In this kind of model (type III), we consider two cases. In the case (a) we
assume the VEV as 
\begin{equation}
\left\langle \Phi _{1}\right\rangle _{0}=\left( 
\begin{array}{c}
0 \\ 
v_{1}/\sqrt{2}
\end{array}
\right) \;\;,\;\;\left\langle \Phi _{2}\right\rangle _{0}=\left( 
\begin{array}{c}
0 \\ 
v_{2}/\sqrt{2}
\end{array}
\right)  \nonumber
\end{equation}
and we take the complex phase of $v_{2}\;$equal to zero since we are not
interested in CP violation. The mass eigenstates of the scalar fields are
given by \cite{moda} 
\begin{eqnarray}
\left( 
\begin{array}{c}
G_{W}^{\pm } \\ 
H^{\pm }
\end{array}
\right) &=&\left( 
\begin{array}{cc}
\cos \beta & \sin \beta \\ 
-\sin \beta & \cos \beta
\end{array}
\right) \left( 
\begin{array}{c}
\phi _{1}^{\pm } \\ 
\phi _{2}^{\pm }
\end{array}
\right) ,  \nonumber \\
\left( 
\begin{array}{c}
G_{Z}^{0} \\ 
A^{0}
\end{array}
\right) &=&\left( 
\begin{array}{cc}
\cos \beta & \sin \beta \\ 
-\sin \beta & \cos \beta
\end{array}
\right) \left( 
\begin{array}{c}
\sqrt{2}Im\phi _{1}^{0} \\ 
\sqrt{2}Im\phi _{2}^{0}
\end{array}
\right) ,  \nonumber \\
\left( 
\begin{array}{c}
H^{0} \\ 
h^{0}
\end{array}
\right) &=&\left( 
\begin{array}{cc}
\cos \alpha & \sin \alpha \\ 
-\sin \alpha & \cos \alpha
\end{array}
\right) \left( 
\begin{array}{c}
\sqrt{2}Re\phi _{1}^{0}-v_{1} \\ 
\sqrt{2}Re\phi _{2}^{0}-v_{2}
\end{array}
\right)  \label{Autoestados masa Higgs}
\end{eqnarray}
where $\tan \beta =v_{2}/v_{1}\;$and $\alpha \;$is the mixing angle of the
CP-even neutral Higgs sector. $G_{Z(W)}\;$are the would-be Goldstone bosons
for $Z\left( W\right) $, respectively. And $A^{0}\;$is the CP-odd neutral
Higgs. $H^{\pm }\;$are the charged physical Higgses.

The case (b) corresponds to the case in which the VEV are taken as 
\begin{equation}
\left\langle \Phi _{1}\right\rangle _{0}=\left( 
\begin{array}{c}
0 \\ 
v_{1}/\sqrt{2}
\end{array}
\right) \;\;,\;\;\ \left\langle \Phi _{2}\right\rangle _{0}=\left( 
\begin{array}{c}
0 \\ 
0
\end{array}
\right) .
\end{equation}
The mass eigenstates scalar fields in this case are \cite{modb} 
\begin{eqnarray}
G_{W}^{\pm } &=&\phi _{1}^{\pm }\;\;\;\;\;\;,\;\;\;\;\;\;H^{\pm }=\phi
_{2}^{\pm }\;\;,  \nonumber \\
G_{Z}^{0} &=&\sqrt{2}Im\phi _{1}^{0}\;\;\;\;,\;\;\;\;A^{0}=\sqrt{2}Im\phi
_{2}^{0},  \label{Autoestados modelo III}
\end{eqnarray}
and the neutral CP-even fields are the same as in the former model just
replacing $v_{2}=0.\;$A very important difference between both models is
that $G_{Z\left( W\right) }\;$is a linear combination of components of $\Phi
_{1}\;$and $\Phi _{2}$ in the model (a), meanwhile in the model (b) is a
component of the doublet $\Phi _{1}.$

\section{Generation of models type I and II from type III}

To convert the lagrangian (\ref{Yukawa}) into mass eigenstates we make the
unitary transformations 
\begin{eqnarray}
D_{L,R} &=&\left( V_{L,R}\right) D_{L,R}^{0}\;  \label{Down transf} \\
U_{L,R} &=&\left( T_{L,R}\right) U_{L,R}^{0}\;  \label{Up transf}
\end{eqnarray}
$\;$

from which we obtain the mass matrices. In the framework of case (a) 
\begin{eqnarray}
M_{D}^{diag} &=&V_{L}\left[ \frac{v_{1}}{\sqrt{2}}\eta ^{D,0}+\frac{v_{2}}{%
\sqrt{2}}\xi ^{D,0}\right] V_{R}^{\dagger }  \label{Masa down} \\
M_{U}^{diag} &=&T_{L}\left[ \frac{v_{1}}{\sqrt{2}}\eta ^{U,0}+\frac{v_{2}}{%
\sqrt{2}}\xi ^{U,0}\right] T_{R}^{\dagger }  \label{Masa up}
\end{eqnarray}

From (\ref{Masa down}), (\ref{Masa up}) we can solve for $\xi ^{D,0},\xi
^{U,0}\;$obtaining

\begin{eqnarray}
\xi ^{D,0} &=&\frac{\sqrt{2}}{v_{2}}V_{L}^{\dagger }M_{D}^{diag}V_{R}-\frac{%
v_{1}}{v_{2}}\eta ^{D,0}  \label{Rotation Id} \\
\xi ^{U,0} &=&\frac{\sqrt{2}}{v_{2}}T_{L}^{\dagger }M_{D}^{diag}T_{R}-\frac{%
v_{1}}{v_{2}}\eta ^{U,0}  \label{Rotation Iu}
\end{eqnarray}

Let us call the eqs (\ref{Rotation Id}), (\ref{Rotation Iu}), rotations of
type I, replacing them into (\ref{Yukawa}) the expanded Lagrangian for up
and down sectors are

\begin{widetext}
\begin{eqnarray}
-\pounds _{Y\left( u\right) }^{\left( a,I\right) } &=&-\frac{g\cot \beta }{%
M_{W}}\overline{U}M_{U}^{diag}KP_{L}DH^{+}-\frac{g}{M_{W}}\overline{U}%
M_{U}^{diag}KP_{L}DG^{+}  \nonumber \\
&&+\frac{g}{\sqrt{2}M_{W}\sin \beta }\overline{U}M_{U}^{diag}U\left( \sin
\alpha H^{0}+\cos \alpha h^{0}\right)  \nonumber \\
&&-\frac{ig}{\sqrt{2}M_{W}}\overline{U}M_{U}^{diag}\gamma _{5}UG^{0}-\frac{%
ig\cot \beta }{\sqrt{2}M_{W}}\overline{U}M_{U}^{diag}\gamma _{5}UA^{0} 
\nonumber \\
&&+\frac{1}{\sin \beta }\overline{U}\eta ^{U}KP_{L}DH^{+}-\frac{1}{\sqrt{2}%
\sin \beta }\overline{U}\eta ^{U}U\left[ \sin \left( \alpha -\beta \right)
H^{0}+\cos \left( \alpha -\beta \right) h^{0}\right]  \nonumber \\
&&+\frac{i}{\sqrt{2}\sin \beta }\overline{U}\eta ^{U}\gamma _{5}UA^{0}+h.c.+%
\text{leptonic\ sector}.  \label{Yukawa 1au}
\end{eqnarray}

\begin{eqnarray}
-\pounds _{Y\left( d\right) }^{\left( a,I\right) } &=&\frac{g\cot \beta }{%
M_{W}}\overline{U}KM_{D}^{diag}P_{R}DH^{+}+\frac{g}{M_{W}}\overline{U}%
KM_{D}^{diag}P_{R}DG^{+}  \nonumber \\
&&+\frac{g}{\sqrt{2}M_{W}\sin \beta }\overline{D}M_{D}^{diag}D\left( \sin
\alpha H^{0}+\cos \alpha h^{0}\right)  \nonumber \\
&&+\frac{ig}{\sqrt{2}M_{W}}\overline{D}M_{D}^{diag}\gamma _{5}DG^{0}+\frac{%
ig\cot \beta }{\sqrt{2}M_{W}}\overline{D}M_{D}^{diag}\gamma _{5}DA^{0} 
\nonumber \\
&&-\frac{1}{\sin \beta }\overline{U}K\eta ^{D}P_{R}DH^{+}-\frac{1}{\sqrt{2}%
\sin \beta }\overline{D}\eta ^{D}D\left[ \sin \left( \alpha -\beta \right)
H^{0}+\cos \left( \alpha -\beta \right) h^{0}\right]  \nonumber \\
&&-\frac{i}{\sqrt{2}\sin \beta }\overline{D}\eta ^{D}\gamma _{5}DA^{0}+h.c.+%
\text{leptonic\ sector.}  \label{Yukawa 1ad}
\end{eqnarray}
\end{widetext}
where $K\;$is the CKM matrix. The superindex $(a,I)\;$refers to the case (a)
and rotation type I.

It is easy to check that if we add (\ref{Yukawa 1au}) and (\ref{Yukawa 1ad})
we obtain a lagrangian consisting of the one in the 2HDM type I\ \cite{moda}%
, plus the FC interactions. Therefore, we obtain the lagrangian of type I
from eqs (\ref{Yukawa 1au}) and (\ref{Yukawa 1ad}) by setting $\eta
^{D}=\eta ^{U}$ $=0.\;$In addition, it is observed that the case (b) in both
up and down sectors can be calculated just taking the limit $\tan \beta
\rightarrow \infty $.

On the other hand, from (\ref{Masa down}), (\ref{Masa up}) we can also solve
for $\eta ^{D,0},\eta ^{U,0}\;$instead of, to get 
\begin{eqnarray}
\eta ^{D,0} &=&\frac{\sqrt{2}}{v_{1}}V_{L}^{\dagger }M_{D}^{diag}V_{R}-\frac{%
v_{2}}{v_{1}}\xi ^{D,0}  \label{Rotation IId} \\
\eta ^{U,0} &=&\frac{\sqrt{2}}{v_{1}}T_{L}^{\dagger }M_{U}^{diag}T_{R}-\frac{%
v_{2}}{v_{1}}\xi ^{U,0}  \label{Rotation IIu}
\end{eqnarray}
which we call rotations of type II, replacing them into (\ref{Yukawa}) the
expanded lagrangian for up and down sectors become 
\begin{widetext}
\begin{eqnarray}
-\pounds _{Y\left( u\right) }^{\left( a,II\right) } &=&\frac{g}{M_{W}}\tan
\beta \overline{U}M_{U}^{diag}KP_{L}DH^{+}-\frac{g}{M_{W}}\overline{U}%
M_{U}^{diag}KP_{L}DG^{+}  \nonumber \\
&&+\frac{g}{\sqrt{2}M_{W}\cos \beta }\overline{U}M_{U}^{diag}U\left( \cos
\alpha H^{0}-\sin \alpha h^{0}\right) -\frac{ig}{\sqrt{2}M_{W}}\overline{U}%
M_{U}^{diag}\gamma _{5}UG^{0}  \nonumber \\
&&+\frac{ig\tan \beta }{\sqrt{2}M_{W}}\overline{U}M_{U}^{diag}\gamma
_{5}UA^{0}-\frac{1}{\cos \beta }\overline{U}\xi ^{U}KP_{L}DH^{+}  \nonumber
\\
&&+\frac{1}{\sqrt{2}\cos \beta }\overline{U}\xi ^{U}U\left[ \sin \left(
\alpha -\beta \right) H^{0}+\cos \left( \alpha -\beta \right) h^{0}\right] -%
\frac{i}{\sqrt{2}\cos \beta }\overline{U}\xi ^{U}\gamma _{5}UA^{0}  \nonumber
\\
&&+h.c.+\text{leptonic sector}  \label{Yukawa 2au}
\end{eqnarray}

\begin{eqnarray}
-\pounds _{Y(d)}^{(a,II)} &=&-\frac{g\tan \beta }{M_{W}}\overline{U}%
KM_{D}^{diag}P_{R}DH^{+}+\frac{g}{M_{W}}\overline{U}KM_{D}^{diag}P_{R}DG^{+}
\nonumber \\
&&+\frac{g}{\sqrt{2}M_{W}\cos \beta }\overline{D}M_{D}^{diag}D\left( \cos
\alpha H^{0}-\sin \alpha h^{0}\right) +\frac{ig}{\sqrt{2}M_{W}}\overline{D}%
M_{D}^{diag}\gamma _{5}DG^{0}  \nonumber \\
&&-\frac{ig\tan \beta }{\sqrt{2}M_{W}}\overline{D}M_{D}^{diag}\gamma
_{5}DA^{0}+\frac{1}{\cos \beta }\overline{U}K\xi ^{D}P_{R}DH^{+}  \nonumber
\\
&&+\frac{1}{\sqrt{2}\cos \beta }\overline{D}\xi ^{D}D\left[ \sin \left(
\alpha -\beta \right) H^{0}+\cos \left( \alpha -\beta \right) h^{0}\right] +%
\frac{i}{\sqrt{2}\cos \beta }\overline{D}\xi ^{D}\gamma _{5}DA^{0}  \nonumber
\\
&&+h.c.+\;\text{leptonic sector}  \label{Yukawa 2ad}
\end{eqnarray}

\end{widetext}

\begin{figure}[h] 
\begin{center} 
\includegraphics[angle=0, width=6cm]{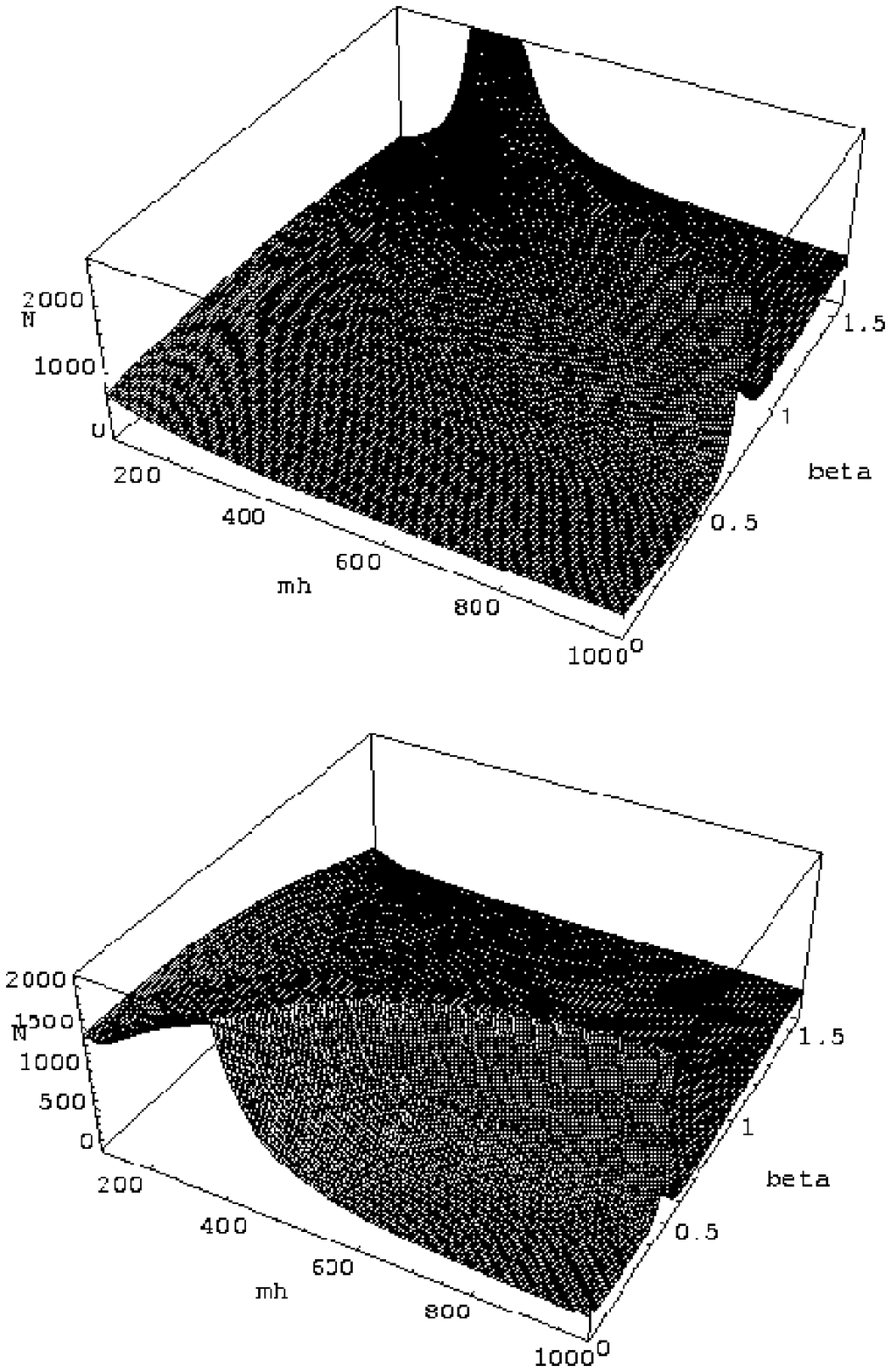} 
\end{center} 
\caption{The figure 1 corresponds to 3D plots of the fraction of FC 
couplings coming from the ratio of the muon contribution and tau 
contribution in the radiative corrections for the process $\protect\mu \to e  
\protect\gamma$. With $\protect\alpha =\protect\pi /16$, $m_{H^0}=300$ GeV 
and $m_{A^{0}}$ is decoupled. The figure on the top corresponds to (aI) and
the other one to (aII).}  
\label{Fig. 1} 
\end{figure} 

In this situation the case (b) is obtained in the limit $\tan \beta
\rightarrow 0,\;$for up and down sectors. Moreover, if we add the
lagrangians (\ref{Yukawa 1au}) and (\ref{Yukawa 2ad}) we find the lagrangian
of the 2HDM type II \cite{moda} plus the FC interactions. Similarly like
before, lagrangian type II is obtained setting $\xi ^{D}=\eta ^{U}=0.$\
Therefore, lagrangian type II is generated by making a rotation of type I in
the up sector and a rotation of type II in the down sector, it is valid
since $\xi ^{U}\;$and $\xi ^{D}\;$are independent each other and same to $%
\eta ^{U,D}.\;$In addition, we can build two additional lagrangians by
adding $\pounds _{Y(u)}^{(a,II)}+\pounds _{Y(d)}^{(a,II)}\;$and $\pounds
_{Y(u)}^{(a,II)}+\pounds _{Y(d)}^{(a,I)}$. So four models are generated from
the case (a). On the other hand, interactions involving Goldstone bosons are
the same in all the models in the R-gauge, while in the unitary gauge they
vanish \cite{moda}.

Finally, we can realize that in both models (a) and (b) with both types of
rotations FCNC processes vanishes when all Higgses are decoupled, we shall
prove it by using the rare processes $\mu \rightarrow eee\;$and $\mu
\rightarrow e\gamma $.

\section{LFV processes}

In the present work, we study the processes $\;\mu \rightarrow e\gamma $ and 
$\mu \rightarrow eee\;$ in the 2HDM type III. The decay width of $\mu
\rightarrow e\gamma \;$in both models (a) and (b) comes from one loop
corrections, where we have used a muon running in the loop. The first
interaction vertex is proportional to the muon mass and the final vertex is
proportional to the flavor changing transition $\mu \rightarrow e.$\ The
decay widths in the two types of rotations are given by

\begin{widetext}
\begin{eqnarray}
\Gamma ^{(a,I)}\left( \mu \rightarrow e\gamma \right) &=&\frac{4G_{F}\alpha
m_{\mu }^{7}\eta _{\mu e}^{2}}{\sqrt{2}\sin ^{4}\beta }\left| \sin \alpha
\sin \left( \alpha -\beta \right) F_{1}(m_{H^{0}}) +\cos \alpha \cos \left( \alpha -\beta \right)
F_{1}(m_{h^{0}})-\cos \beta F_{2}(m_{A^{0}})\right| ^{2}  \nonumber \\
\Gamma ^{\left( a,II\right) }\left( \mu \rightarrow e\gamma \right) &=&\frac{%
4G_{F}\alpha m_{\mu }^{7}\xi _{\mu e}^{2}}{\sqrt{2}\cos ^{4}\beta }\left|
-\cos \alpha \sin \left( \alpha -\beta \right) F_{1}(m_{H^{0}})
 +\sin \alpha \cos \left( \alpha -\beta \right)
F_{1}(m_{h^{0}})-\tan \beta F_{2}(m_{A^{0}})\right| ^{2}
\end{eqnarray}
\end{widetext}
where 
\begin{eqnarray}
F_{1}(x) &=&\frac{\log [x^{2}/m_{\mu }^{2}]}{4\pi ^{2}x^{2}}  \nonumber \\
F_{2}(x) &=&\frac{-\log \left[ x^{2}/m_{\mu }^{2}\right] }{8\pi ^{2}x^{2}}
\end{eqnarray}

\begin{figure}[h]
\begin{center}
\includegraphics[angle=0, width=6cm]{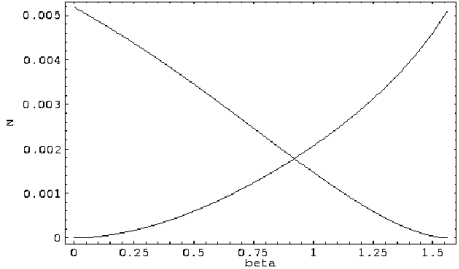}
\end{center}
\caption{The figure 2 illustrate the differences between the models (aI) and
(aII) respect to the parameter $\tan \protect\beta$. We have decoupled the
higgses masses $m_{H^0}=300$ GeV and $m_{A^{0}}$ and taken $\protect\alpha =%
\protect\pi /16$ and $m_{h^0=}300$ GeV. The curve that increases with $\beta$
corresponds to the model (aI).}
\label{Fig. 2}
\end{figure}

The decay widths for the process $\mu \rightarrow eee\;$ in the two cases
read 

\begin{eqnarray}
\Gamma ^{\left( a,I\right) }\left( \mu \rightarrow eee\right) &=&\frac{
2G_{F}m_{\mu }^{5}m_{e}^{2}N_{\mu e}^{2}}{\sqrt{2}1024\pi ^{3}\sin ^{4}\beta 
}\left| \frac{\sin \alpha \sin \left( \alpha -\beta \right)
}{m_{H^{0}}^{2}}\right. \nonumber \\ 
&+&
\left. \frac{\cos \alpha \cos \left( \alpha -\beta
\right)}{m_{h^{0}}^{2}}-\frac{\cos \beta }{m_{A^{0}}^{2}}\right| ^{2}, 
\nonumber\\
 \Gamma ^{\left( a,II\right) }\left( \mu \rightarrow eee\right)
&=&\frac{
 2G_{F}m_{\mu }^{5}m_{e}^{2}N_{\mu e}^{2}}{\sqrt{2}1024\pi ^{3}\cos
^{4}\beta 
 }\left| \frac{\cos \alpha \sin \left( \alpha -\beta \right)
}{m_{H^{0}}^{2}} \right.\nonumber \\ 
&-&\left.
 \frac{\sin \alpha \cos \left( \alpha
-\beta \right)}{m_{h^{0}}^{2}} +\frac{ \sin \beta }{m_{A^{0}}^{2}}\right|
^{2},
  \end{eqnarray}

And the corresponding expresions for the case (b) are obtained taking the
appropiate limits. These FC processes vanish when all Higgses are decoupled.

Now, by using the experimental upper bounds for LFV processes \cite{bolton,
bell} 
\begin{eqnarray}
\Gamma \left( \mu \rightarrow e\gamma \right) &\leq &3.59\times
10^{-30}\;GeV,  \nonumber \\
\Gamma \left( \mu \rightarrow eee\right) &\leq &3.0\times 10^{-31}\;GeV,
\end{eqnarray}

We see that the upper bounds imposed by $\mu \rightarrow e\gamma \;$are much
more restrictive.

We use a muon running in the loop for the calculation of $\mu \rightarrow
e\gamma \;$instead of a tau as customary. This would be reasonable provided
some conditions. If we take the quotient $\Gamma ^{\left( a,\tau \right) }/$ 
$\Gamma ^{\left( a,\mu \right) }\;$where $\Gamma ^{\left( a,\mu \right) }\;$
represents the width of $\mu \rightarrow e\gamma \;$with a muon in the loop
for the case (a), and similarly for $\Gamma ^{\left( a,\tau \right) }\;$,
and we set $m_{H^{0}}=300$ GeV, $\alpha =\pi /16$ and $m_{A}$ is decoupled,
we can plot the quotient 
\begin{equation}
\frac{N_{\mu e}}{N_{\mu \tau }N_{\tau e}}
\end{equation}
by supposing that $\Gamma ^{(a,\mu )}\approx \Gamma ^{(a,\tau )}$, i. e.,
they are of the same order. Here $N_{\mu e}\;$ denotes the FC coupling in a
generic way. We can notice from figure 1 that the values obtained
 for the
fraction cover a wide range and therefore this assumption is
 reasonable.

 We turn now to derive constraints for
arbitrary values of the Higgs
sector. Let us consider the process $\mu
 \rightarrow e\gamma \;$in both cases
for different values of the Higgs
 masses and mixing angles. In the figure 2
we take $m_{h^{0}}$ and $m_{A^{0}}\;$ going to infinity. We plot $N_{\mu
e}\;vs\;\beta \;$,  for $\alpha =\pi /16\;$
 and $m_{h^{0}}=300$ GeV for the
models $aI(aII)\;$ respectively. We can
 observe that the behaviour of the
models are quite different in a long range
 of $\tan \beta$. Additionally,
near to the critical points of $\tan \beta \;$
 the models take complementary
values.

\begin{figure}[h]
\begin{center}
\includegraphics[angle=0, width=6cm]{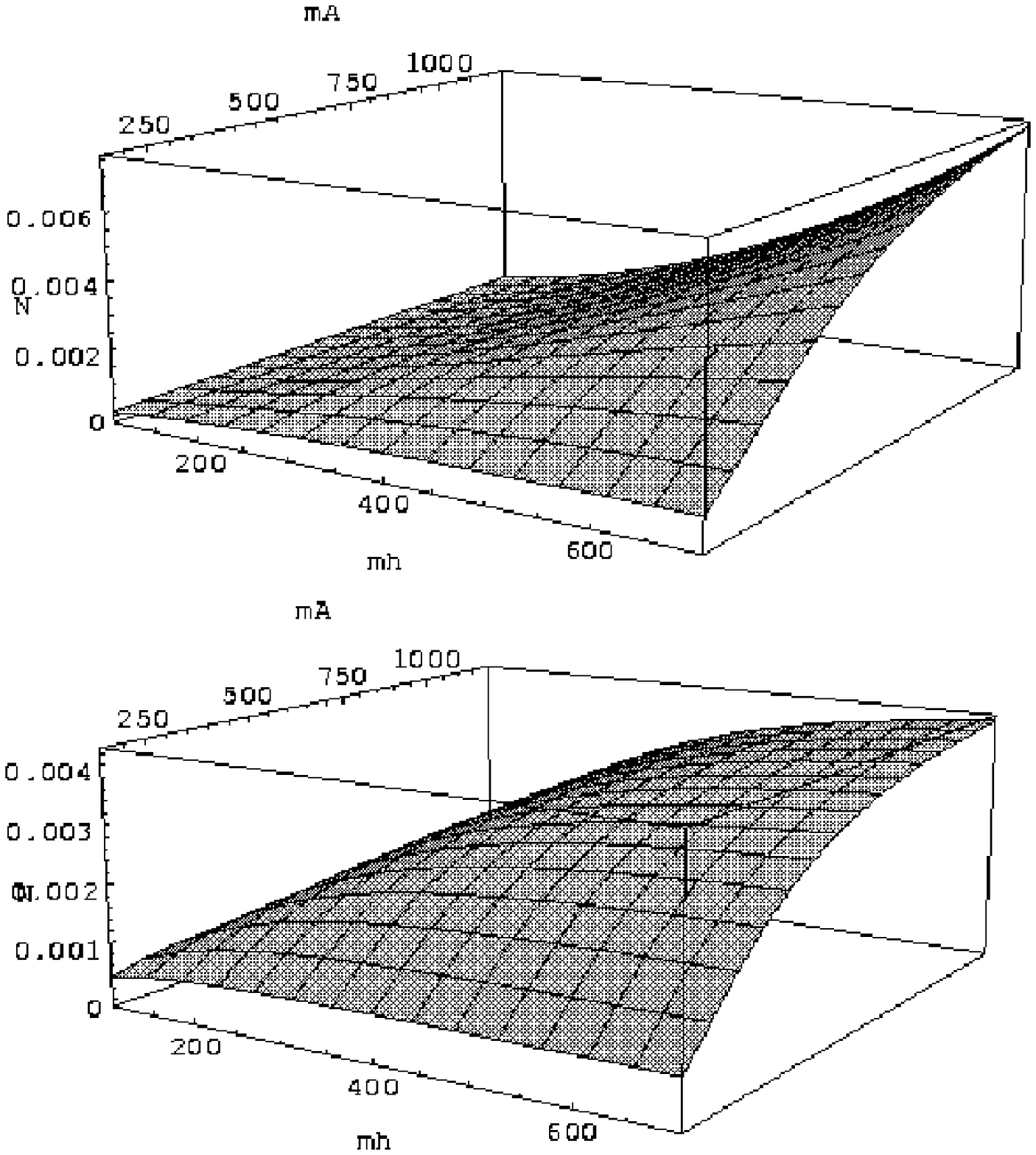}
\end{center}
\caption{The figure 3 is for the parameter space $(N_{\protect\mu e}, m_h,
m_A)$ for the models (aI) and (aII) respectively. We set $\tan
\protect%
 \beta=1$,the higgs mass $m_{H^0}=500$ GeV and $\protect\alpha
=\protect\pi
 /16$.}
\label{Fig. 3}
\end{figure}
 
The 3D plots $(N_{\mu e}, m_h, m_A)$ are shown in the figure 3 for $m_H=500$
GeV, $\alpha= \pi/16$ and $\tan \beta=1$. They represent the models (aI) and
(aII), similar to the figure 2. Once again, we realize that the behaviour of
both models is quite different.

The figure 4 corresponds to the models (aII) and (bII) in which $m_{H^{0}}=300$
 GeV and $\alpha =\pi /16.\;$ For the model (aII) we use $\tan
\beta =1.\;$ These graphics illustrate that the cases (a) and (b) are
substantially different.

\begin{figure}[h]
\begin{center}
\includegraphics[angle=0, width=6cm]{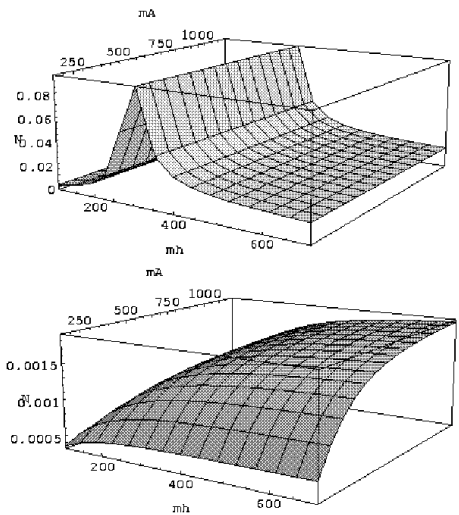}
\end{center}
\caption{Figure 4 shows the differences between models (a) and (b), it is
plotted in the
 parameter space $(N_{\protect\mu e}, m_h, m_A)$. The parameter
$\tan \protect
 \beta=1$ for the model (a),the higgs mass $m_{H^0}=300$
GeV and $\protect\alpha =\protect\pi /16$.} 
\label{Fig. 4}
\end{figure}

\section{Conclusions}

In the present work we examine a 2HDM type III which produces FCNC at tree
level in the leptonic sector. We classified the model type III according to
the VEV taken by the Higgses and to the method used to rotate the mixing
matrices. All that, in order to write down the lagrangian in the mass
eigenstates. When both doublets acquire a VEV we talk about the case (a),
while when only one doublet acquire a VEV we talk about the case (b). On the
other hand, when we write $\xi ^{D,0},\xi ^{U,0}\;$ in terms of $\eta
^{D,0},\eta ^{U,0}\;$ plus the mass matrices, it is called here a rotation
of type I. Where $\xi ^{D,0},\xi ^{U,0}\;$ are the mixing matrices which
couple to  $\Phi _{2}\;$ and $\widetilde{\Phi }_{2}\;$ respectively and
 $\eta ^{D,0},\eta ^{U,0}\;$ are the FC matrices which couple to  $\Phi
_{1}\;$and $\widetilde{\Phi }_{1}\;$ respectively. Now, when we solve for 
$\eta ^{D,0},\eta ^{U,0}\;$ in terms of $\xi ^{D,0},\xi ^{U,0}\;$ and the
mass matrices we call it a rotation of type II.

In addition, we observe that the 2HDM of type I plus FC interactions is
generated by adding the lagrangian of type (a,I) in the up sector and the
lagrangian of type (a,I) in the down sector, meanwhile the lagrangian of
type II plus FC interactions is generated by adding the lagrangian of type
(a,I) in the up sector and the lagrangian of type (a,II) in the down sector.
Other two combinations are possible i.e. $\pounds _{Y\left( u\right)
}^{\left( a,II\right) }+\pounds _{Y\left( d\right)}^{\left( a,I\right)}$ and 
$\pounds _{Y\left( u\right) }^{\left(a,II\right)} +\pounds_{Y\left(
d\right)}^{\left( a,II\right) }$. Moreover, if we began with a lagrangian of
type (a,I) we would obtain the lagrangian (b,I) taking the limit $\tan \beta
\rightarrow \infty ,\;$ while if we started with a lagrangian of type (a,II)
we would obtain the lagrangian (b,II) in the limit $\tan \beta \rightarrow 0$.

To illustrate the importance of this classification we show graphics to find
bounds on the FC coupling $N_{\mu e}\;$ coming from the process $\mu
\rightarrow e\gamma \;$and we realize that such bounds are sensitive to the
type of rotation and also to the structure of the VEV. We also calculate the
process $\mu \to 3e$ for both kind of rotations but the constraints obtained
were
 less restrictive than the ones obtained with the process $\mu \to e
\gamma$.
 
Finally, to evaluate such bounds we have used a muon running in the loop
for
 the process $\mu \rightarrow e\gamma \;$instead of a tau as usual.
Consequently,  we plot the
quotient $N_{\mu
 e}/(N_{\mu \tau }N_{e\tau })\;$in terms of $m_{h^{0}}\;$and
$\beta \;$, getting a wide range of allowed values for that quotient,
showing that this assumption is reasonable.

We acknowledge to M. Nowakowski for his suggestions and for the careful
reading of the manuscript. This work was supported by COLCIENCIAS.

\end{document}